\documentclass[twocolumn,showpacs,preprintnumbers,amsmath,amssymb,showkeys]{revtex4-1}
\usepackage{graphicx}
\usepackage{caption}
\usepackage{pstricks}
\usepackage[dvips]{hyperref}
\usepackage[TS1,OT1,T1]{fontenc}
\usepackage{dsfont}
\usepackage{float}

\providecommand{\abs}[1]{\lvert#1\rvert}

\providecommand{\Fig}[1]{Figure~\ref{#1}}
\newcommand{\dps}{\displaystyle}
\newcommand{\VT}{V_{\rm tot}}

\begin{document}
\title{Local density dependent potential for compressible mesoparticles}

\author{G\'er\^ome Faure}
\author{Jean-Bernard Maillet}
\affiliation{CEA, DAM, DIF, F-91297 Arpajon, France}

\author{Gabriel Stoltz}
\affiliation{Universit\'e Paris-Est, CERMICS (ENPC), INRIA, F-77455 Marne-la-Vall\'ee, France}

\date{\today}


\begin{abstract}
We focus on finding a coarse grained description able to reproduce
the thermodynamic behavior of a molecular system by using
mesoparticles representing several molecules. Interactions between
mesoparticles are modelled by an interparticle potential, and an
additional internal equation of state is used to account for the
thermic contribution of coarse grained internal degrees of
freedom. Moreover, as strong non-equilibrium situations over a wide
range of pressure and density are targeted, the internal
compressibility of these mesoparticles has to be considered. This is
done by introducing a dependence of the potential on the local
environment of the mesoparticles, either by defining a spherical local
density or by means of a Voronoi tessellation. As an example, a local density dependent potential is fitted to reproduce the Hugoniot curve of a model of nitromethane,
where each mesoparticle represents one thousand molecules.
\end{abstract}
\maketitle

\section{Introduction}
Since two decades, the development of coarse graining strategies from
all-atoms classical molecular dynamics (MD) to the mesoscale has known a continuous growing
interest. This is particularly relevant for complex systems where longer time and length scales behaviors should be adressed. A first
step toward the development of reduced models was the idea of united
atoms potentials, where some atoms belonging to the same
molecule are considered as a single center of force, without any
modification of the equations of motion (EoM). A general framework for
coarse graining which employs modified EoM appeared with the
Dissipative Particle Dynamics (DPD) method~\cite{hoogerbrugge92}, where a particular mesoscopic model can be deduced from its atomistic representation~\cite{flekkoy_2000}. In this method, coarse grained
degrees of freedom are introduced, the effect of the lost degrees of freedom being 
modelled by the addition of dissipative and stochastic terms in the
EoM. A relation between the amplitude of fluctuations and dissipation guaranties that the canonical distribution is sampled~\cite{espanol_1995}. DPD has finally become a standard method to simulate complex fluids at the mesoscale, and is generally associated with soft potentials allowing to use larger integration timesteps. However, for target simulations in the microcanonical
ensemble (NVE) or for non-equilibrium situations, an additional variable
has to be introduced to guaranty the conservation of the total 
energy~\cite{espanol_1997, avalos97, strachan05, stoltz_2006}, leading to DPD with conserved energy (DPDE). This is particularly useful in
the case of shock wave simulations, where the equilibrium temperature
in the shocked state relies on the ability of the molecule to store
energy, as the kinetic energy is split between intermolecular
(i.e. center of mass motion) and intramolecular (vibrons) motions. For
complex large molecules, the intramolecular part is the dominant
term. The additional variable, the internal energy of the mesoparticle,
is linked to the internal temperature through an internal equation of
state ($\epsilon=\int C_v(T)dT$). The heat capacity $C_v(T)$ represents the thermic
contribution of the internal coarse-grained degrees of freedom. DPDE
allows to reproduce thermodynamic properties of molecular
systems over a wide range of pressures and densities, as well as
non-equilibrium shock wave situations. One hidden approximation during
this coarse graining of a molecule to a mesoparticle is the implicit
hypothesis of the rigidity of the molecule, i.e., the molecule is not
compressible at all, and only thermic effects have been considered so
far. This hypothesis is no longer valid when several molecules are represented by a single
mesoparticle: the compressibility
of this ensemble of molecule is obviously not null (and eventually
tends to the compressibility of the whole system when mesoparticles
get larger). Such "internal" compressibility can be accounted for by introducing a dependence of the
parameters of the effective interaction potential on the local
density.

Density dependent potentials (DDP) have been widely used in the literature, aiming originally at modelling specific effects as for example the bonding directionality for covalent materials with the Tersoff potentials \cite{tersoff86}, or metallic bonding using the embedded atom method \cite{daw84}. One has to distinguish between global and local density dependent potential, and be aware that there is not a unique way to fit their parameters \cite{louis_2002}. Nevertheless, DDP have received renewed attention in the last few years, particularly to assess the question of transferability of the potential and to gain accuracy in the prediction of thermodynamic properties\cite{merabia_2007, izvekov_2010, izvekov_2011}. The main concern here is the transferability of the potential over a wide range of pressures and densities. To this aim, we present in this article a model of compressible mesoparticles using local density dependent potentials. 

\bigskip

\subsection*{Outline}

This article is organized as follows. We first recall in
Section~\ref{section:local_density} two approaches to defining a local
density: either through some local spherical averages
(Section~\ref{section:spherical_density}) or using Voronoi cells to define
a local volume (Section~\ref{subsection:Voronoi}). It turns out that both
approaches lead to very different results, as carefully documented in
Section~\ref{subsection:comparison}. We therefore favor the Voronoi approach
since it is, as we argue, more transferable with respect to
modifications of the local thermodynamic conditions. In a second step,
we describe in Section~\ref{section:density_dep_pot} the density dependent
potentials we use, and discuss their tendency to (strongly) alter the
average pressure in the system. Finally, as a stringent application,
we reproduce the Hugoniot curve of a nitromethane like material in
Section~\ref{section:Hugoniot}.

\section{Defining a local density}
\label{section:local_density}

\subsection{Locally averaged density}
\label{section:spherical_density}

Consider an ensemble of $N$~(meso)particles of mass $m$ located at
positions $\{ q_i \}_{i=1,\dots,N}$. The standard approach to define a
continuous density field from this discrete set of masses is to use
weight functions centered around particle centers. We consider a
smooth, non negative, spherically symmetric function~$\chi$ vanishing
for $r \geq r_{\rm cut}$. Denoting by $r_{i,j} = |q_i-q_j|$ the
distance between two particles, the local density $\rho_i^{\rm s}$ of
particle $i$ reads
\begin{equation}
\rho_i^s = \frac{1}{4\pi} \frac{\dps \sum_{j=1}^N m\chi(r_{i,j})}{\dps \int_0^{r_{\rm cut}} \chi(r)\, r^2 dr}.
\label{dstloc_def}
\end{equation}
We consider in the sequel three examples of weighting function:
\begin{itemize}
\item a third order spline function $\chi_{\rm sp}$ such that
  $\chi_{\rm sp}(0) = 1$, $\chi_{\rm sp}(r_{\rm cut}) = 0$ and
  $\chi_{\rm sp}'(0) = \chi_{\rm sp}'(r_{\rm cut}) = 0$;
\item a smoothed step function $\chi_{\rm ssf}$ such that $\chi_{\rm
  ssf}(r) = 1$ for $r<0.9\,r_{\rm cut}$, which is smoothly going down
  to 0 for $0.9 \, r_{\rm cut} \leq r \leq r_{\rm cut}$ (thanks to a
  cubic spline);
\item the Lucy function, commonly used in SPH models \cite{lucy77}
\[
\dps \chi_{\rm lf}(r) = \left(1+3\frac{r}{r_{\rm cut}}\right) \left(1-\frac{r}{r_{\rm cut}}\right)^3.
\]
\end{itemize}

\begin{figure}[H]
\begin{center}
\rotatebox{270}{\includegraphics[scale=0.32]{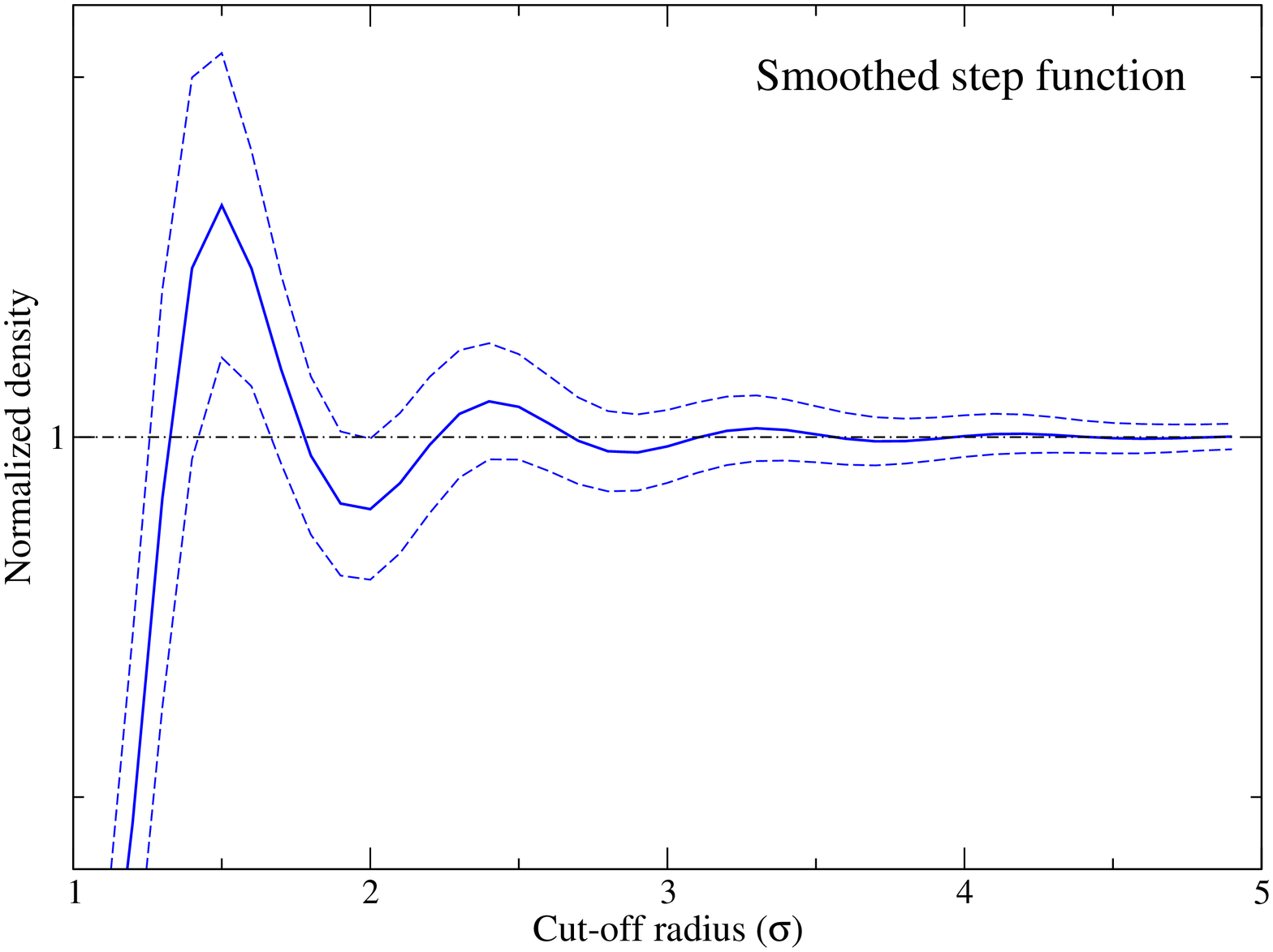}}
\rotatebox{270}{\includegraphics[scale=0.32]{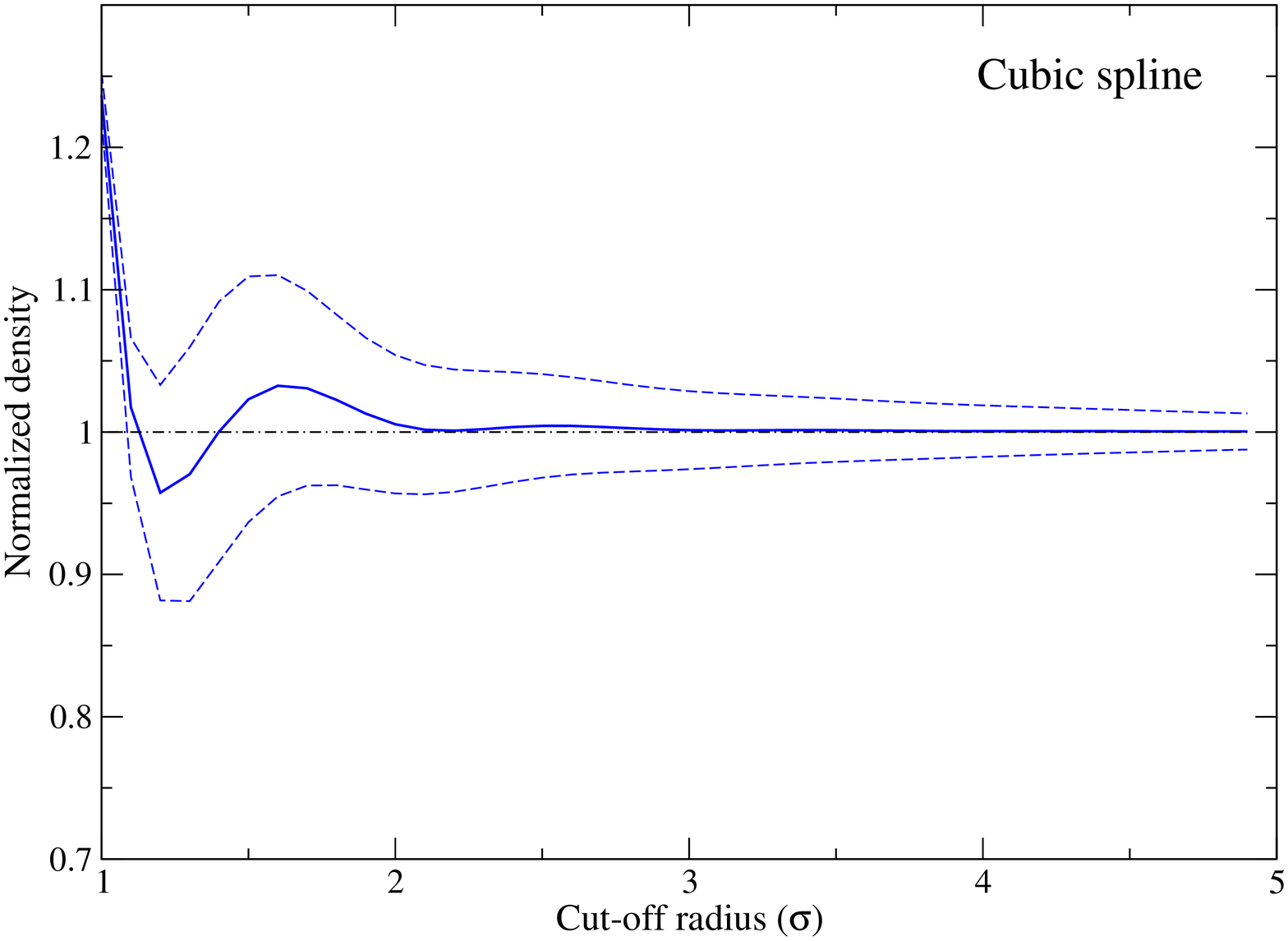}}
\rotatebox{270}{\includegraphics[scale=0.32]{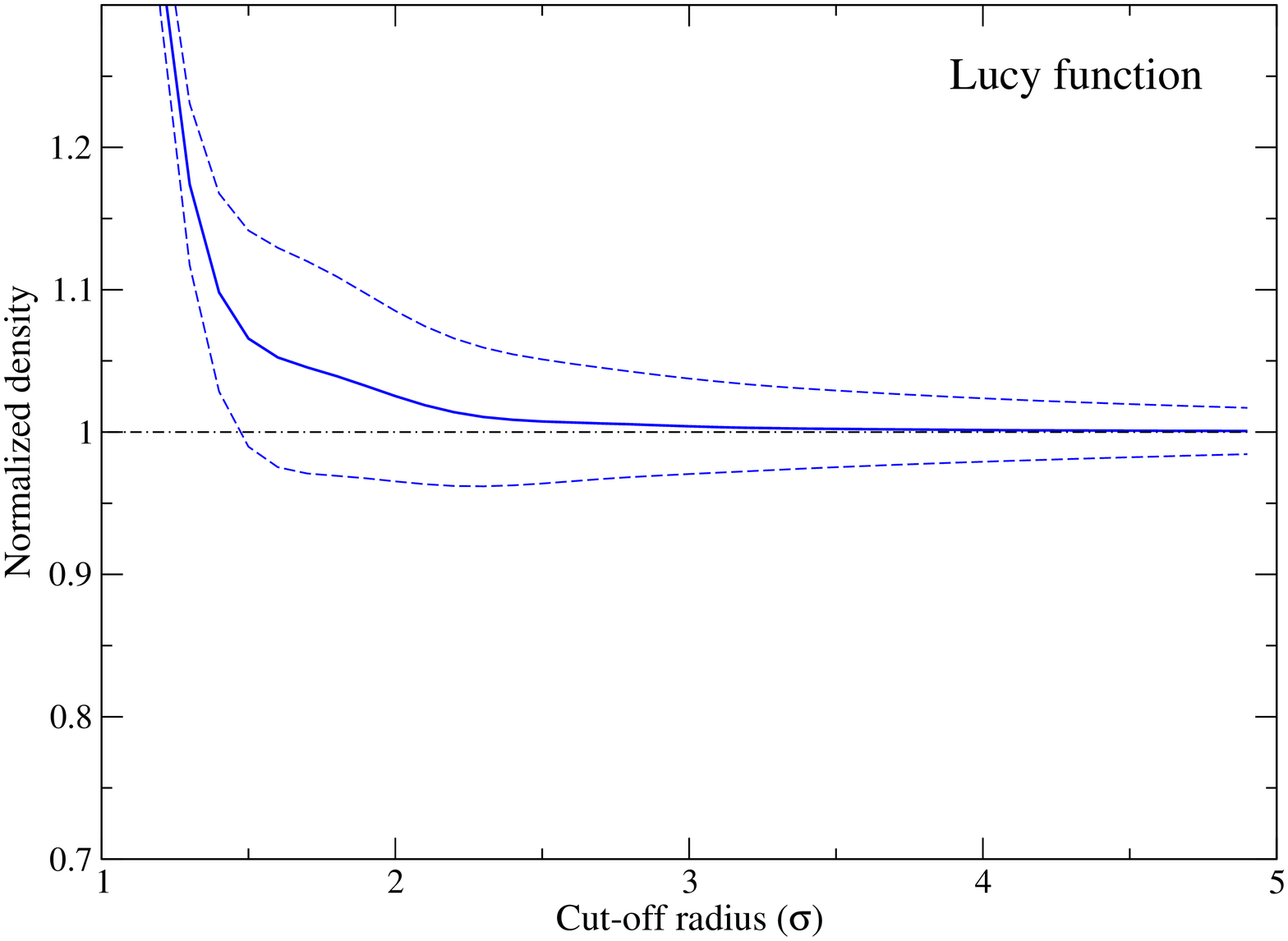}}
\caption{Mean local density $\frac{\langle \rho^{\rm s}
    \rangle}{\rho_{\rm th}}$ (solid line), and the envelop
  corresponding to the standard deviation of the distributions of densities (dashed lines) with respect
  to the cut-off radius $r_{\rm cut}$. Top: smoothed step function
  $\chi_{\rm ssf}$. Middle: cubic spline $\chi_{\rm sp}$. Bottom: Lucy
  function $\chi_{\rm lf}$.}
\label{dstloc_rcut}
\end{center}
\end{figure}

The definition of the local densities crucially depends on the choice
of the weight function and of its cut-off radius. The mean local
density $\langle \rho^{\rm s} \rangle$, computed by averaging the
local densities $\{ \rho_i^{\rm s} \}_{i=1,\dots,N}$ using the
arithmetic average
\begin{equation}
\label{eq:arithmetic_average}
\langle \rho^{\rm s} \rangle = \frac1N \sum_{i=1}^N \rho_i^{\rm s},
\end{equation}
is plotted as a function of the cut-off radius $r_{\rm cut}$ for the
previously defined weight functions in~\Fig{dstloc_rcut}. The
simulation is performed for Argon at temperature $T = 300$~K and
density $\rho_{\rm th} = 1650$~kg.m$^{-3}$. Argon is described by a
Lennard-Jones potential with parameters $\sigma = 3.4$~{\AA} and
$\varepsilon/k_{\rm B} = 120$~K, truncated at $r_{\rm LJ} = 2.5 \,
\sigma$. In all this work, constant temperature simulations are
performed in the NVT ensemble, which is sampled with a Langevin
dynamics (the integrator we use and the parameters of the dynamics are
described in Appendix~\ref{section:langevin}).

\Fig{dstloc_rcut} shows that the thermodynamic density $\rho_{\rm th}$
and the mean local density $\langle \rho^{\rm s} \rangle$ agree for
large cut-off radii, as expected since local environment effects are
averaged out. On the other hand, too small a radius causes large
discrepancies between $\rho_{\rm th}$ and $\langle \rho^{\rm s}
\rangle$ due to the structure of the radial distribution
function. This suggests to consider large radii in order to avoid
systematic biases in the simulation. However, this option is
expensive, and suppresses any information on the density variations.

We have checked that other averages, for instance, harmonic averages,
also lead to biased mean spherical densities (see
Appendix~\ref{section:other_mean}).

The bias in the mean local density is often removed by renormalizing
the local densities as $\rho_i^{\rm s}\frac{\rho_{\rm th}}{\langle
  \rho^{\rm s} \rangle}$, or by computing the theoretical mean local
density at $r_{\rm cut}$ given an a priori equation of
state~\cite{almarza_2003}. These remedies however hide the actual
issue at stake, namely model inconsistency. In any case, local density
averages cannot be applied to the simulation of shock compressions
where the density is a priori unknown and strongly inhomogeneous in
non-equilibrium systems.

\subsection{Voronoi tessellation}
\label{subsection:Voronoi}

An alternative way to define local densities is to introduce a local
notion of volume associated with a particle. In order to recover the
thermodynamic density in average, we use a partitioning of the total
volume using the particle centers as reference points to construct a
Voronoi tessellation. This is a standard approach in mesoscale models
of dissipative particle dynamics~\cite{serrano_2002}. However, the
equations governing the evolution of these models are of hydrodynamic
type, with postulated equations of states.  In contrast, the model we
consider in this article is still of atomistic nature, with particles
interacting through potential energy functions.

\begin{figure}[!ht]
\begin{center}
\includegraphics[scale=0.32]{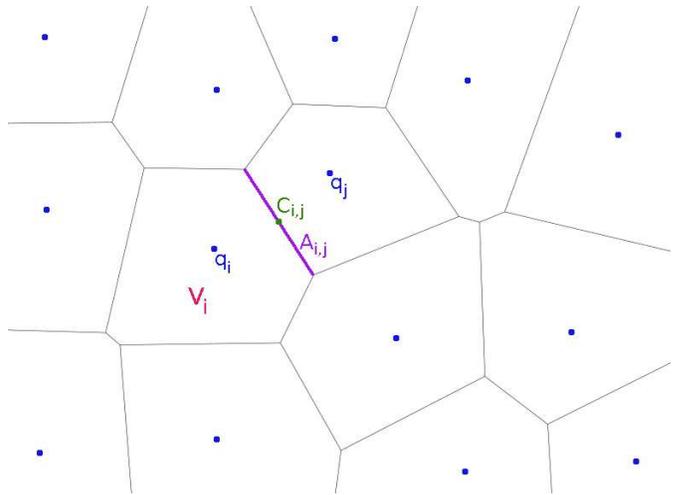}
\end{center}
\caption{\label{voro_ex}
A 2D Voronoi tessellation. Two cells $R_i,R_j$, of respective centers
$\mathbf{q}_i, \mathbf{q}_j$ and volumes $V_i,V_j$ are
highlighted. The area of the surface separating these cells is denoted
$A_{i,j}$, $\mathbf{C}_{i,j}$ being the centroid of the corresponding
face.}
\end{figure}

We denote by $R_i$ the Voronoi cell associated with the $i$th particle
(defined as the set of points closer to $q_i$ than any other particle
$q_j$, see~\Fig{voro_ex}), and by $V_i = |R_i|$ its volume. The local
density is defined as
\[
\rho_i^{\rm v} = \frac{m}{V_i},
\]
while the average density~$\overline{\rho^{\rm v}}$ in the system is defined through the harmonic average
\[
\frac{1}{\overline{\rho^{\rm v}}} = \frac1N \sum_{i=1}^N \frac{1}{\rho_i^{\rm v}}.
\]
Since the Voronoi tessellation defines a partition of the space (of
total volume $\VT$), we automatically have $\sum V_i = \VT$, thus
$\overline{\rho^{\rm v}} = \rho_{\rm th}$. Besides, the Voronoi volume
does not depend on a weight function or a parameter such as a cut-off
radius, which is a clear advantage to unambiguously define a local
density.

Since we aim at performing molecular dynamics with interaction potentials depending on the local density, hence on the Voronoi volumes, the derivatives of the volumes $V_i$ with respect to the particle positions are needed. As shown in~\cite{serrano_2001}, it holds, for $j \neq i$,
\begin{equation}
  \nabla_{\mathbf{q}_j}V_i = -\frac{A_{i,j}}{r_{i,j}}(\mathbf{C}_{i,j}-\mathbf{q}_j),
  \label{voro_der}
\end{equation}
where $A_{i,j}$ and $\mathbf{C}_{i,j}$ are respectively the area and
the centroid of the face of the Voronoi diagram between cells $i$ and
$j$. The derivative with respect to the position of particle $i$ can
easily be inferred from the other derivatives given
in~\eqref{voro_der} by using the invariance of the Voronoi volume with
respect to a translation, which implies
\begin{equation}
\sum_j \nabla_{\mathbf{q}_j} V_i = 0.
\label{voro_der2}
\end{equation}

Let us also recall an interesting property of the Voronoi volumes
under dilation or compression, which is useful to understand average
pressures in materials described by density dependent potentials (see
Section~\ref{subsection:pot_anvir}):
\begin{equation}
\sum_j (\nabla_{\mathbf{q}_j} V_i)\cdot \mathbf{q}_j = 3V_i.
\label{voro_der3}
\end{equation}

\subsection{Comparison of the spherical and Voronoi densities}
\label{subsection:comparison}

We now investigate the relationship between the two notions of local
density defined in Sections~\ref{section:spherical_density} and
\ref{subsection:Voronoi}. We expect particles with high spherical local
densities to have small Voronoi volumes. On the other hand, since the
local density $\rho_i^{\rm s}$ is computed from the positions of all
the particles within a certain radius, it is unclear whether it agrees
with the Voronoi densities, which depend only on the nearest
neighbors.

\begin{figure}[!ht]
\begin{center}
\rotatebox{270}{\includegraphics[scale=0.32]{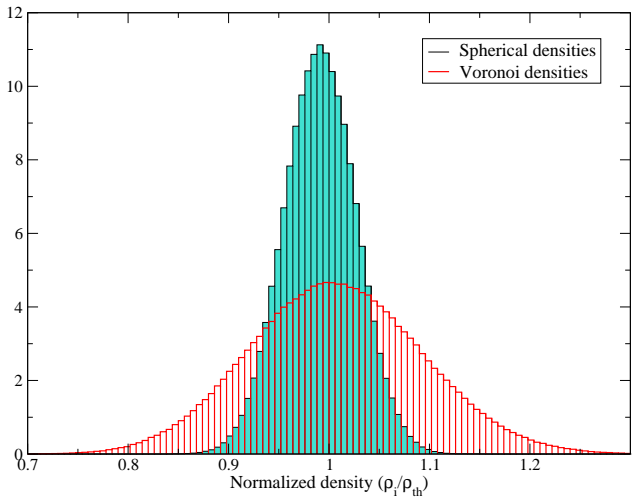}}
\end{center}
\caption{Distribution of the spherical ($r_{\rm cut} = 2.5\sigma$) and Voronoi densities at $\rho = 1650$~kg.m$^{-3}$}
\label{dst_hist}
\end{figure}

Figure~\ref{dst_hist} presents a histogram of the Voronoi and
spherical densities (with $r_{\rm cut} = r_{\rm LJ}$ for the latter
one) in the NVT ensemble at $T=300$~K and $\rho_{\rm th} =
1650$~kg.m$^{-3}$, using a Lennard-Jones potential for Argon. Both can
reasonnably be considered as Gaussian. We find here a standard
deviation of $0.0854\rho_{\rm th}$ for the Voronoi density. The
standard deviation of the spherical density depends on the cut-off
radius as Figure~\ref{dstloc_rcut} suggests. At $r_{\rm cut} = 2.5 \,
\sigma$, it is smaller than the variance of the Voronoi volume and
equal to $0.0359 \rho_{\rm th}$.

In order to more quantitatively compare the two
notions of densities, we next compute the correlation between
normalized densities. We introduce to this end the standard deviation
for the spherical densities using the arithmetic
average~\eqref{eq:arithmetic_average}:
\[
\mathrm{sd}(\rho^{\rm s}) =\dps \sqrt{\frac1N \sum_{i=1}^N (\rho_i^{\rm s}-\langle\rho^{\rm s} \rangle)^2},
\]
and similarly for the Voronoi densities
\[
\mathrm{sd}(\rho^{\rm v}) =\dps \sqrt{\frac1N \sum_{i=1}^N (\rho_i^{\rm v}-\langle\rho^{\rm v} \rangle)^2}.
\]
We then consider the normaized densities $\tilde{\rho}_i^{\rm s}$ and $\tilde{\rho}_i^{\rm v}$, which have mean~$0$ and variance~$1$, namely
\[
\tilde{\rho}_i^{\rm s} = \dps \frac{\rho_i^{\rm s} - \langle \rho^{\rm s} \rangle}{\mathrm{sd}(\rho^{\rm s})}, \qquad
\tilde{\rho}_i^{\rm v} = \dps \frac{\rho_i^{\rm v} - \langle \rho^{\rm v} \rangle}{\mathrm{sd}(\rho^{\rm v})},
\]
and compute the correlation
\begin{equation}
C = \dps \frac1N \sum_{i=1}^N \tilde{\rho}_i^{\rm s}\tilde{\rho}_i^{\rm v}.
\label{def_correl}
\end{equation}

\begin{figure}[!ht]
\begin{center}
\includegraphics[angle=270,scale=0.32]{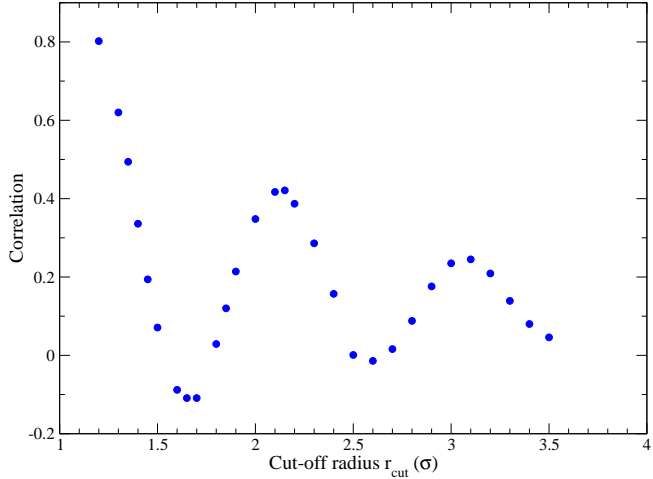}
\caption{Correlation~\eqref{def_correl} between the Voronoi and local densities with respect to the local density cut-off radius~$r_{\rm cut}$.}
\label{dstloc_voro_correl}
\end{center}
\end{figure}

As demonstrated by~\Fig{dstloc_voro_correl}, the correlation is quite
high when the cut-off radius of the spherical averages is small
(\textit{i.e.} when the same neighbors are taken into account). It
however varies in a non-monotonic way, first decreasing until $r_{\rm
  cut} \simeq 1.7\sigma$, then increasing again until $r_{\rm cut}
\simeq 2.2\,\sigma$, and still oscillating for larger radii. In any
case, the correlation remains quite small.

\begin{figure}[!ht]
\begin{center}
\includegraphics[angle=270,scale=0.32]{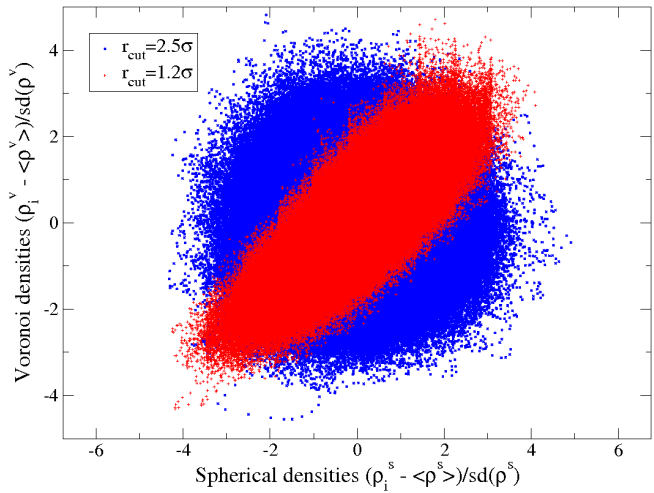}
\caption{Joint distribution of the Voronoi and spherical densities at $r_{\rm cut} = 1.2 \, \sigma$ (red) and $r_{\rm cut} = 2.5 \, \sigma$ (blue).}
\label{dstloc_voro_jd}
\end{center}
\end{figure}

We investigate more precisely the relationship at the important values
$r_{\rm cut} = 2.5\,\sigma$ (which is also the potential energy cutoff
$r_{\rm LJ}$) and $r_{\rm cut} = 1.2\,\sigma$ (where the correlation
is maximum), by plotting in~\Fig{dstloc_voro_jd} the joint
distribution of $(\tilde{\rho}_i^{\rm s},\tilde{\rho}_i^{\rm v})$. The
correlation in the case $r_{\rm cut} = 2.5\,\sigma$ is close to~0,
which is indicated by the absence of trend in the joint distribution,
in sharp contrast with the joint distribution at $r_{\rm cut} =
1.2\,\sigma$. 

\section{Introducing a local density dependence in the potential}
\label{section:density_dep_pot}

We present in this section a possible density dependent potential to
be used in conjunction with a definition of local densities based on a
Voronoi tessellation. We carefully study the average pressure computed
from the corresponding physical model. Let us indeed insist that, to
properly simulate and predict the behavior of materials under shock
compressions, the equation of state needs to be correctly described.

\subsection{Definition of a local density dependent potential}
\label{sec:def_local_density_dep_pot}

We present in this section a way to incorporate local density effects
in the interaction potential. We consider systems with pairwise
forces, such as exp-6 or Lennard-Jones, and denote the interaction
energy between two particles~$i$ and~$j$ by $\mathcal{U}_{\rm
  std}(r_{i,j})$. One possible choice is to correct the distance
$r_{i,j}$ between the particle centers to account for local density
effects, by replacing $r_{i,j}$ by $r_{i,j} - \lambda_{i,j}$ with
\[
\lambda_{i,j} = \lambda(V_{i,j}), \qquad V_{i,j} = \frac{V_i+V_j}{2}.
\]
The associated interaction energy (density dependent potential $U_{\rm dd}$) reads
\begin{equation}
\mathcal{U}_{\rm dd}(r_{i,j},V_{i,j}) = \mathcal{U_{\rm std}}\left( r_{i,j} - \lambda_{i,j}\right),
\label{dd_def}
\end{equation}
so that the total potential energy is
\[
\mathcal{U}_{\rm tot}(q_1,\dots,q_N) = \sum_{1 \leq i < j \leq N} \mathcal{U}\left( r_{i,j} - \lambda_{i,j}\right).
\]
The physical idea underpinning the choice of the correction
$\lambda_{i,j}$ is that the interactions between mesoparticles arise
from the individual interactions of the atoms or molecules represented
by the mesoparticles. This interaction should be dominated by atoms or
molecules on the surface of the mesoparticles. We therefore need to
use $r_{i,j}-\lambda_0$ as the interaction distance at some density
$\rho_0$, with $\lambda_0 \geq 0$ representing the size of the
mesoparticle. When the density increases, the distance between the
centers of the mesoparticles decreases but their sizes should also decrease to mimic their internal compressibility. Hence the correction $\lambda$, i.e. the mesoparticle diameter, should decrease as the density increases (or equivalently $\lambda$ should be an increasing function of the voronoi volume).

\begin{figure}[!ht]
\begin{center}
\includegraphics[scale=0.32]{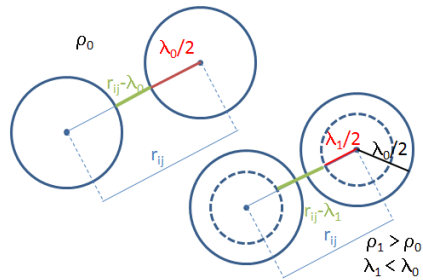}
\end{center}
\caption{\label{fig:dist_corr}Correction of the density effect on the interaction distance.}
\end{figure}

In fact, this can be seen as a way of correcting the interaction distance to take into
account the compressibility of the mesoparticles to prevent the system
from becoming too stiff (see Figure~\ref{fig:dist_corr}).

Another way to introduce a density dependence is discussed in
Appendix~\ref{section:other_dd_pot}. The approach described there
essentially consists in assigning effective volumes to the particles
depending on the local density.

Although obvious, it is worth emphasizing that the dependence of the
volumes $V_i,V_j$ on the particle positions induces extra terms in the
expressions of the atomic forces as compared to the forces arising
from the standard potential $\mathcal{U}_{\rm std}$. More precisely,
\begin{equation}
\begin{aligned}
& \mathbf{F}_{\rm dd}^{i,j,k} = -\nabla_{\mathbf{q}_k} \mathcal{U}_{\rm dd}(r_{i,j},V_{i,j}) \\
& =\left( (\delta_{k,i}-\delta_{k,j})\mathbf{e}_{i,j} + \lambda'(V_{i,j})\nabla_{\mathbf{q}_k}V_{i,j} \right) \mathcal{U}_{\rm std}'\left(r_{i,j}-\lambda_{i,j}\right)
\end{aligned}
\label{dd_force}
\end{equation}
where $\mathbf{e}_{i,j} = (q_j-q_i)/r_{i,j}$.

\subsection{Analytic calculation of the virial pressure for density dependent potentials}
\label{subsection:pot_anvir}

We show in this section that the pressure computed with density
dependent potentials may be very different from the one computed with
standard potentials. We focus on the potential part of the pressure,
by considering the so-called virial pressure
\[
\begin{aligned}
P_{\rm pot}(q)& = -\frac{1}{3 \VT}\sum_{k=1}^N \mathbf{q}_k \cdot \nabla_{\mathbf{q}_k} \mathcal{U}_{\rm tot} = \frac{1}{3 \VT}\sum_{1 \leq i < j \leq N} w_{\rm dd}^{i,j},
\end{aligned}
\]
with
\[
w_{\rm dd}^{i,j} = -\sum_{k=1}^N \mathbf{q}_k \cdot \nabla_{\mathbf{q}_k}\mathcal{U}\left( r_{i,j} - \lambda_{i,j}\right).
\]
This expression should be compared to the contribution arising only from the direct pairwise interaction between~$i$ and~$j$:
\[
w_{\rm direct}^{i,j} = -r_{i,j} \, \mathcal{U}'_{\rm std}(r_{i,j}-\lambda_{i,j}).
\]
The expression of the force~\eqref{dd_force} shows that $w_{\rm dd}^{i,j}$ is related to $w_{\rm direct}^{i,j}$ as:
\begin{equation}
w_{\rm dd}^{i,j} =  w_{\rm direct}^{i,j}\!\!\left(\!1 - \!\sum_{k=1}^N \frac{\lambda'(V_{i,j})}{r_{i,j}}\nabla_{\mathbf{q}_k}\!\!\left(\frac{V_i+V_j}{2}\right)\!\cdot\mathbf{q}_k \right).
\label{dd_pr}
\nonumber
\end{equation}
Using~\eqref{voro_der3}, we finally obtain:
\begin{equation}
w_{\rm dd}^{i,j} = w_{\rm direct}^{i,j}\left(1 - 3\frac{\lambda'(V_{i,j})V_{i,j}}{r_{i,j}}\right).
\label{vp_dd_voro}
\end{equation}
The variations of $\lambda_{i,j}$ induce modifications to the global
pressure compared to the case when $\lambda$ is independent of
the positions of the particles. As shown by the numerical experiments
presented in Section~\ref{subsection:obs_press}, these modifications may be
significant. As all thermodynamic and structural properties cannot be maintained accurately during the coarse graining process \cite{johnson07}, we focus in the sequel on the pressure and specifically optimize the coarse grained potential to reproduce the pressure of the reference system.

\subsection{Numerical observations}
\label{subsection:obs_press}

We model the dependence of $\lambda_{i,j}$ as a function of the Voronoi volumes for instance as 
\begin{equation}
\lambda_{i,j} = S\left(\frac{V_{i,j}}{V_0}-1\right),
\label{vp_voro}
\end{equation}
where $S$ is a real parameter which can be positive or negative, and
$V_0$ is the average volume per particle $\VT/N$. The reference
situation corresponds to $S=0$, in which case there is no density
dependence in the interaction potentials. Note that~\eqref{vp_voro}
ensures that the (arithmetic) average value $\langle \lambda \rangle$
is always~0, whatever the value of~$S$. 

Our aim is to systematically study the average potential pressure as a
function of the parameter $S$ in~\eqref{vp_voro}. The Voronoi
tessellation was implemented in 3D using the C++ library
Voro++~\cite{rycroft_2009} with periodic boundary conditions. We use
the potential from~\cite{maillet_2011} as the standard potential $\mathcal{U}_{\rm std}$, \textit{i.e,} an
Exponential-6 potential :
\begin{equation}
  \mathcal{U}_{\rm std}(r) = \frac{\varepsilon}{\alpha-6} \left( 6\exp\left(\alpha\left[1-\frac{r}{\sigma}\right]\right) - \alpha \left(\frac{\sigma}{r}\right)^6\right).
  \label{exp6_nm}
\end{equation}
The parameters set to $\sigma = 4.5$~\AA,
$\varepsilon/k_{\rm B} = 427$~K and $\alpha = 26$. The values of these
parameters are motivated by coarse-grained models of nitromethane
where one particle stands for one molecule. The simulations were
carried out in the NVT ensemble at room temperature ($T=300$~K) and at a
density chosen such that the average pressure vanishes ($\rho_{\rm th}
= 1144$~kg.m$^3$).

\begin{figure}[!ht]
\begin{center}
\includegraphics[angle=270,scale=0.32]{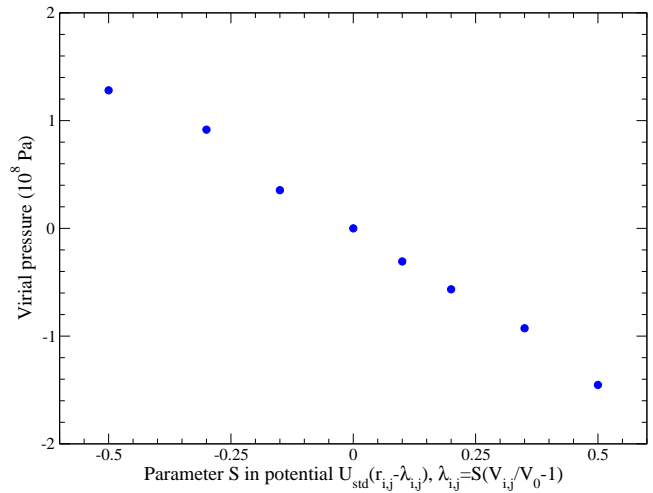}
\caption{Virial pressure computed for the Voronoi - dependent Exp-6 potential. The standard case corresponds to $S=0$.}
\label{pvoro_pr}
\end{center}
\end{figure}

Whereas the kinetic pressure is of course the same in all these
simulations, we plot in Figure~\ref{pvoro_pr} the total potential
pressure for various values of~$S$. The results show that genuinely
density dependent potentials ($S\neq0$) lead to strong deviations of
the potential pressure compared to the case $S=0$. The sign of the
deviation depends on the sign of~$S$, in accordance
with~\eqref{vp_dd_voro} since, in view of~\eqref{vp_voro},
\[
w_{\rm dd}^{i,j} = w_{\rm direct}^{i,j}\left(1 - 3\frac{S V_{i,j}}{V_0 r_{i,j}}\right).
\]
The virial pressure is therefore increased when $S < 0$, and decreased when $S > 0$. Indeed, $S > 0$ means that the diameter of the mesoparticle (i.e. proportional to $\lambda$) decreases as its Voronoi volume decreases (or as its local density increases), hence lowering the forces (and the pressure) between the mesoparticles. This is the expected physical behavior of the mesoparticle. On the contrary, $S < 0$ means that $\lambda$ is a decreasing function of the voronoi volume, meaning that the diameter of the mesoparticles increases as the density increases (hence increasing the pressure). This is an unwanted effect for physical applications.

\section{Hugoniot curve for a mesoparticle model of nitromethane}
\label{section:Hugoniot}

We now turn to our main application, namely the computation of
physical properties under strong regimes of pressure and temperature
for models of mesoparticles. This is indeed a stringent test of the
quality of the density dependent potentials suggested in
Section~\ref{section:density_dep_pot}.

More precisely, we compute thermodynamic states along the Hugoniot
curve for a model of mesoparticle representing 1,000 molecules of
nitromethane (CH$_3$NO$_2$). The molar mass of the mesoparticules is therefore~61 kg/mol. The potential energy function we use is the same as in
Section~\ref{subsection:obs_press}, except that we change the parameters of the
potential to $\sigma = 9$~\AA, $\varepsilon/k_{\rm B} = 300$~K and
$\alpha = 17$. These parameters are obtained by fitting low density part of the Hugoniot curve. Reference data for Hugoniot curves are taken from the
previous all-atom Monte Carlo simulations using a united atom
model~\cite{desbiens_2009}.

The important point to correctly reproduce the reference data is the
functional form of the distance correction~$\lambda(V_{i,j})$. We
propose here a two step optimization method. We first consider in
Section~\ref{subsection:suggested} a potential independent of the density,
\textit{i.e.} $\lambda(V_{i,j}) = \lambda_0$ fixed. We determine the
value of $\lambda_0$ which reproduces the reference pressure $P_{\rm
  th}$ at each given thermodynamic density $\rho_{\rm th}$. In a
second step, we propose in Section~\ref{subsection:optimized} to fit the
so-computed values of $\lambda_0$ by linear or quadratic functions,
and optimize upon the parameters in the fitting functions to fully
reproduce the Hugoniot curve.

\subsection{Suggested functional form of the distance correction}
\label{subsection:suggested}

In order to have some data on how the distance correction $\lambda$
may depend on the thermodynamic conditions, we first start by fitting
its value on a series of $N_{\rm ts}$ thermodynamic states along the
reference Hugoniot curve. We label these states by $s$ and index them
by their densities $\rho_s$ and temperatures $T_s$. The optimal value
of $\lambda_0$ for each thermodynamic condition is denoted by
$\lambda_0(\rho_s,T_s)$ and is chosen to minimize the error in
pressure $E = \abs{P_{\rm sim}(\rho_s,T_s)-P_{\mathrm{ref},s}}$
independently for each density $\rho_s$. We use here a non-dependent
potential taking a constant value of $\lambda_0$ in~\eqref{exp6_nm}
and run simulations in the NVT ensemble during an integration time
$t_{\rm sim} = 1$~ns to obtain $P_{\rm sim}(\rho_s,T_s)$. The pressures resulting from this first
optimization, displayed as orange squares in Figure~\ref{hug_tot}
and labeled as 'independent optimization', are by construction almost on top of reference data .

\begin{figure}[!ht]
\begin{center}
\includegraphics[angle=270,scale=0.32]{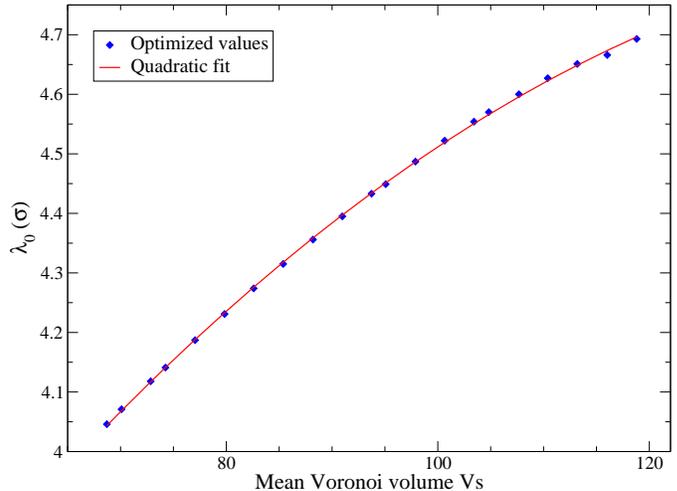}
\caption{Optimal $\lambda_0$ for the reference states with respect to the mean Voronoi volume $V_s = \frac{m_{\rm sys}}{\rho_{s}}$}
\label{r0_opt}
\end{center}
\end{figure}

Figure~\ref{r0_opt} presents the values of $\lambda_0$ as a function
of $V_s = \frac{m_{\rm sys}}{\rho_s}$. This suggests to interpolate
the values of $r_0$ by a polynomial fit. A quadratic least-square fit
is displayed in Figure~\ref{r0_opt}.

\subsection{Optimized distance correction}
\label{subsection:optimized}

\begin{figure}[!ht]
\begin{center}
\includegraphics[angle=270,scale=0.32]{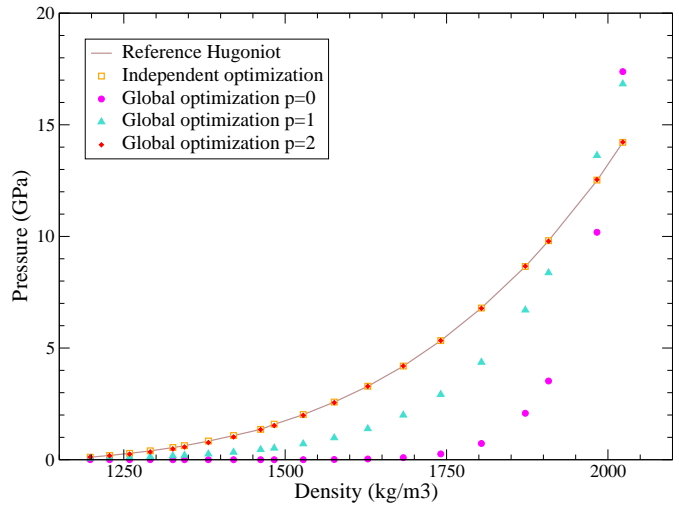}
\end{center}
\caption{Hugoniot curve for nitromethane: reference curve~\cite{desbiens_2009} and results for density-dependent potentials with $p=0$ and $p=2$.}
\label{hug_tot}
\end{figure}

We now fit the distance correction on the whole set of thermodynamic states $(\rho_s,T_s)_{1 \leq s \leq N_{\rm s}}$ by a polynomial function of degree $p$:
\begin{equation}
\lambda(V) = \sum_{i=0}^{p} a_i V^i.
\label{r0_fit}
\end{equation}
The coefficients $\{ a_i \}$ are chosen in order to minimize the mean quadratic error in pressure: 
\[
F(\{ a_i \}) = \sqrt{\frac{1}{N_{\rm ts}} \sum_{s=1}^{N_{\rm ts}} (P_{\rm sim}(\rho_s,T_s) - P_{\mathrm{ref},s})^2}.
\]
In this expression, $P_{\rm sim}(\rho_s,T_s)$ is the average pressure
in the NVT ensemble at density~$\rho_s$ and temperature~$T_s$ for the
potential energy function~\eqref{dd_def} with the distance
correction~\eqref{r0_fit}. We carry out the minimization of this
quadratic error $F$ with a Newton algorithm. The first and second
order derivatives of $F$ are approximated by computing the canonical
average of the observables obtained by differentiating the pressure
observable with respect to the coefficients~$\{ a_i \}$.
The initial guess for the values of $\{ a_i \}$ is obtained by a
least-square fit based on Figure~\ref{r0_opt}. Note that the parameter~$\lambda$ decrease when the density increases 
(or equivalently as the Voronoi volume decrease), which is indeed the expected physical behavior 
discussed at the beginning of Section~\ref{sec:def_local_density_dep_pot}. Consistently with the
observations made in Section~\ref{subsection:obs_press}, directly using this
initial guess leads to poor results since the pressure is largely
underestimated. The
optimization procedure to minimize the mean quadratic error $F(\{ a_i
\})$ is therefore mandatory:
\begin{itemize}

\item We first show that it is impossible to reproduce the correct
  pressure by using the same constant, non-dependent $\lambda_0$ for
  every density $\rho_s$ (which corresponds to $p=0$
  in~\eqref{r0_fit}). The minimization of $F$ yields $a_{0,\rm min} =
  4.0588\,\sigma$ and $F(a_{0,\rm min}) = 3.131$~GPa (filled circles in Figure~\ref{hug_tot}). This is a prohibitively large error, revealing a poor agreement with the reference curve. The brutal increase in pressure at $\rho \approx 1900$~kg.m$^{-3}$ clearly underlines the non-transferability of the non-dependent potential ($p=0$) to a large set of thermodynamic states. This shows that we need to consider genuinely density dependent potentials, namely $p \geq 1$.

\item The minimization of~$F$ as a function of $(a_0,a_1)$ when $p=1$
  gives $a_{0,\rm min} = 3.1886\, \sigma$ and $a_{1,\rm min} =
  1.3420\cdot10^{-2}\,\sigma^{-2}$ with an error $F(a_{0,\rm
    min},a_{1,\rm min}) = 1.396$~GPa. Although lower than with $p=0$,
  such a large error still means the computed Hugoniot curve (blue triangles in Figure~\ref{hug_tot}) largely deviates from the reference curve, with too large a curvature. This suggest to further increase the degree of the polynomial $\lambda(V)$.

\item The minimization of a quadratic distance
correction $\lambda$ ($p=2$) leads to $a_{0, \rm min} =
2.4784\,\sigma$, $a_{1, \rm min} = 3.1095\cdot10^{-2}\,\sigma^{-2}$
and $a_{2,\rm min} = -1.0247\cdot10^{-4}\,\sigma^{-5}$. The
corresponding error $F(a_{0,\rm min},a_{1,\rm min},a_{2,\rm min}) =
0.034$~GPa shows an almost perfect agreement between the reference
Hugoniot and the predicted one (red diamonds in Figure~\ref{hug_tot}).

\end{itemize}

We show in Appendix~\ref{section:predict} that our density dependent
potential provides satisfactory results at even larger densities and
pressures without carrying another optimization to include these
thermodynamic states in the optimization set. This can be seen as some
transferability property.

\section{Conclusion}

We have proposed a method to model the compressibility of
mesoparticles representing a large collection of molecules. This
method ensures that accurate thermodynamics properties (i.e. Equation
of State) can be preserved during the coarse graining process of
molecular systems. As the scale of the modelling increases, two
additional terms, namely the thermic and the cold (pressure)
contributions of coarse grained degrees of freedom, have to be taken
into account for the computation of the total energy. None of these
terms appear in the original DPD framework. The thermic contribution
appeared later, as non-equilibrium situations of mesoparticules were
considered. The second term, the internal compressibility of
mesoparticle, has been implicitly neglected when coarse graining
single molecules, or when the thermodynamic domain of interest
remains near standard conditions. This term becomes significant when
several molecules are embedded in a single mesoparticle, and when a
large range of pressures and densities is targeted.

We explored the possibility of modelling the compressibility of
mesoparticles by introducing a local dependence in our potential. Two
local quantities were studied to that purpose: a spherical local
density within a cut-off radius and a volume defined by a Voronoi
tessellation.
Introducing a dependence of either of these local quantities in the
interaction potential, we noticed a significant increase of pressure
in our simulations. This increase has to be accounted for when fitting
density dependent potentials to reproduce Hugoniot curves for nitromethane
mesoparticles.
Finally, we managed to accurately reproduce the Hugoniot curve of
model nitromethane for mesoparticles containing thousand
molecules. For this resolution, a gain of several order of magnitude
in the CPU time is obtained both from the increase in the
integration timestep and from the lower number of interactions.

\appendix

\section{Langevin dynamics}
\label{section:langevin}
To carry out simulations in the NVT ensemble, we resort to Langevin
dynamics. Given a potential energy function $\mathcal{U}_{\rm tot}(q)$, the equations of motion read 
\[
\begin{aligned}
dq_t &= \frac{p_t}{m}\,dt\\
dp_t &= -\nabla \mathcal{U}_{\rm tot}(q_t)\,dt - \gamma \frac{p_t}{m}\,dt + \sqrt{\frac{2\gamma}{\beta}} \, dW_t,
\end{aligned}
\]
where $\gamma > 0$ is the friction parameter and $W_t$ is a standard
Brownian motion. We discretize here the Langevin dynamics by splitting
the Hamiltonian part of the dynamic and the thermostat part. We use a
velocity-Verlet integration scheme for the Hamiltonian part, and
integrate analytically the thermostat part (see~\cite{LMS13} for a numerical analaysis of the error on the invariant measure). This gives the following scheme:
\[
\left\{ \begin{aligned}
\tilde{\mathbf{p}}_i^{n+1/2} &= \mathbf{p}_i^{n} -\nabla_{\mathbf{q}_i} \mathcal{U}_{\rm tot}(\mathbf{q}^{n})\frac{\Delta t}{2},\\
\mathbf{q}_i^{n+1} &= \mathbf{q}_i^{n} + \frac{\mathbf{p}_i^{n+1/2}}{m}\Delta t,\\
\widetilde{\mathbf{p}}_i^{n+1} &= \mathbf{p}_i^{n+1/2} - \nabla_{\mathbf{q}_i} \mathcal{U}_{\rm tot}(\mathbf{q}^{n+1})\frac{\Delta t}{2},\\
\mathbf{p}_i^{n+1} & = \exp(-\gamma\,\Delta t)\widetilde{\mathbf{p}}_i^{n+1} + \sqrt{\frac{m(1-\exp(-2\gamma \, \Delta t))}{\beta}}\, G_i^n,
\end{aligned} \right.
\]
where $(G_i^n)_{i,n}$ are independent standard normal variables.

In the reduced units defined by the mass $m$ of a particle, the length
scale $\sigma$ and the energy scale $\varepsilon$ in~\eqref{exp6_nm},
the friction parameter is chosen to be $\gamma = 1$, while the time
step is $\Delta t = 0.001$.

\section{Harmonic mean of the spherical local density}
\label{section:other_mean}

\begin{figure}[!ht]
\begin{center}
\includegraphics[angle=270,scale=0.32]{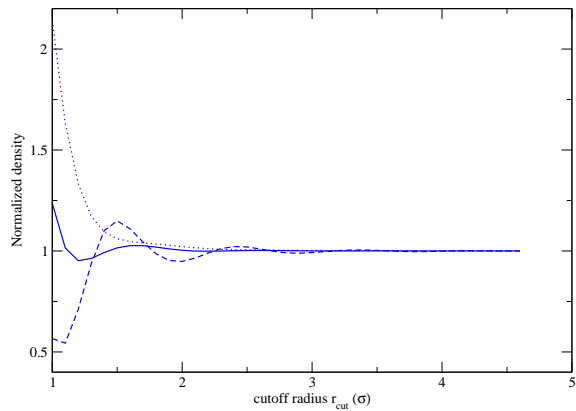}
\caption{Harmonic mean local density for the Lucy function (dotted line), the smoothed step function (dashed line) and the cubic spline (solid line).}
\label{dloc_harm}
\end{center}
\end{figure}

Since we define the average Voronoi density through a harmonic mean, it
is fair to look at the behavior of the spherical densities under harmonic
averages. Figure~\ref{dloc_harm} presents the
evolution of the harmonic mean spherical density as a function of the
cut-off radius for the weight functions studied in
Section~\ref{section:spherical_density}. The harmonic mean clearly
presents the same issues as the arithmetic mean of the spherical
density, namely a bias in the mean spherical density.

\section{Local density dependent Lennard-Jones potential}
\label{section:other_dd_pot}

We explore in this section another general example of density
dependent potential based on a reference potential $\mathcal{U}_{\rm
  std}$. In constrast to~\eqref{dd_def}, it is possible to introduce a
volume dependence by rescaling the distances:
\[
\mathcal{U}_{\rm dd}(r_{i,j},V_{i,j}) = \mathcal{U}_{\rm std}\left( \frac{l_0}{l_{i,j}} r_{i,j}\right),
\]
where $l_{i,j} = l(V_{i,j})$. A relevant example for this kind of dependent potential would be a
Lennard-Jones potential where the length scale $\l$ becomes
dependent on the local density in a way which ensures that $l_0 =
l_{\rm std}$, the length scale for the standard Lennard-Jones
potential.

As in Section~\ref{section:density_dep_pot}, the introduction of a
dependence on the Voronoi volume in the potential naturally causes
extra forces to appear when deriving the potential energy.
\[
\begin{aligned}
\mathbf{F}_{\rm dd}^{i,j,k} & = -\nabla_{\mathbf{q}_k} \mathcal{U}_{\rm dd}(r_{i,j},V_{i,j}) \\
& =\left( (\delta_{k,i}-\delta_{k,j})\mathbf{e}_{i,j} + \frac{l'(V_{i,j})}{l_{i,j}}r_{i,j}\nabla_{\mathbf{q}_k}V_{i,j} \right) \times \\
& \qquad \qquad \qquad \qquad \frac{l_0}{l_{i,j}}\mathcal{U}_{\rm std}'\left(\frac{l_0}{l_{i,j}}r_{i,j}\right). 
\end{aligned}
\]
As in~\eqref{dd_pr}, the contribution to the virial pressure of the interaction between particles $i$ and $j$ is altered by these extra forces:
\begin{align*}
w_{\rm dd}^{i,j} &= \sum_k \mathbf{F}_{\mu}^{i,j,k}\cdot\mathbf{q}_k \notag\\
&= w_{\rm direct}^{i,j}\left(1 - \sum_k \frac{l'(V_{i,j})}{2l_{i,j}}\nabla_{\mathbf{q}_k}(V_i+V_j)\cdot\mathbf{q}_k \right),
\end{align*}
where $w_{\rm direct}^{i,j} =
-\frac{l_0}{l_{i,j}}\mathcal{U}_{\rm
  std}'\left(r_{i,j}\frac{l_0}{l_{i,j}}\right)r_{i,j}$. Using~\eqref{voro_der3},
we finally obtain (compare~\eqref{vp_dd_voro})
\[
w_{\rm dd}^{i,j} = w_{\rm direct}^{i,j}\left(1 - \frac{3l'(V_{i,j})V_{i,j}}{l_{i,j}}\right).
\]

\section{Test of the density dependent potential outside the optimization range}
\label{section:predict}

We study the ability of our density dependent potential to predict the
correct pressure for densities outside the optimization range (here
$1170$~kg.m$^{-3} \leq \rho_s \leq 2023$~kg.m$^{-3}$). We use a
quadratic distance correction $\lambda$ with the parameters optimized
used to produce the results of Figure~\ref{hug_tot}. We test the potential on a new set of
$N_{\rm test}$ thermodynamic states indexed by $k$ ($2100$~kg.m$^{-3}
\leq \rho_{\mathrm{test},k} \leq 2350$~kg.m$^{-3}$) by estimating the relative
mean quadratic error
\[
F_{\rm test}(\{a_i\}) = \sqrt{\frac{1}{N_{\rm test}} \sum_{k=1}^{N_{\rm test}} \frac{(P_{\rm sim}(\rho_{\mathrm{test},k},T_{\mathrm{test},k}) - P_{\mathrm{ref},k})^2}{P_{\mathrm{ref},k}}}.
\] 

\begin{figure}[!ht]
\begin{center}
\includegraphics[angle=270,scale=0.32]{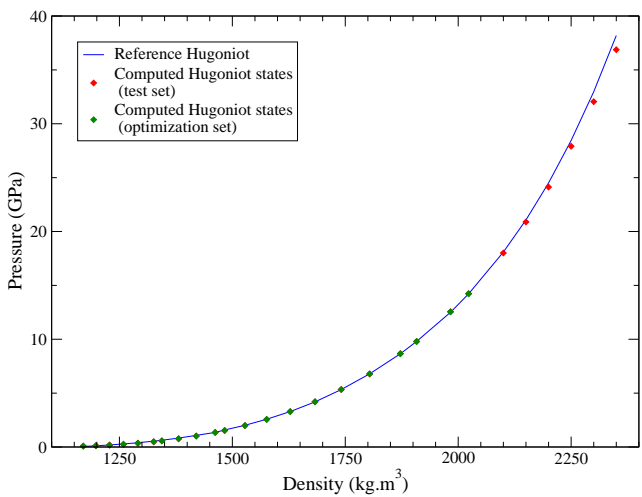}
\end{center}
\caption{Hugoniot curve for nitromethane: Predicted values of the pressure outside the optimization range of the parameters.}
\label{fig:hug_pred}
\end{figure}

The relative error on the test set of thermodynamic states $F_{\rm test} =
0.015$ is  larger than the error on the optimization set
($F = 0.002$), but the computed results are still in excellent agreement with the reference
Hugoniot curve as shown in Figure~\ref{fig:hug_pred}.

\bibliographystyle{biblio}
\bibliography{voro_mp}

\end{document}